# Causal Decomposition in the Mutual Causation System

Albert C. Yang[1], Norden E. Huang[2], Chung-Kang Peng[1]

[1] Division of Interdisciplinary Medicine and Biotechnology, Beth Israel Deaconess Medical Center/Harvard Medical School, Boston, Massachusetts, USA

[2] Center for Dynamical Biomarkers and Translational Medicine, National Central University, Chungli, Taiwan

**Corresponding Author:** Dr. Albert C. Yang, M.D., Ph.D.

Division of Interdisciplinary Medicine and Biotechnology, Beth Israel Deaconess Medical Center/Harvard Medical School

KS-B26, 330 Brookline Ave

Boston, MA, 02215, U.S.A.

Email: cyang1@bidmc.harvard.edu

**Manuscript Information**: 5 figures

**Word Counts:** Abstract (215 words); Main Text (3904 words).

**Abstract**

Inference of causality in time series has been principally based on the prediction paradigm. Nonetheless, the predictive causality approach may overlook the simultaneous and reciprocal nature of causal interactions observed in real world phenomena. Here, we present a causal decomposition approach that is not based on prediction, but based on the instantaneous phase dependency between the intrinsic components of a decomposed time series. The method involves two assumptions: (1) any cause–effect relationship can be quantified with instantaneous phase dependency between the source and target decomposed as intrinsic components at specific time scale, and (2) the phase dynamics in the target originating from the source are separable from the target itself. Using empirical mode decomposition, we show that the causal interaction is encoded in instantaneous phase dependency at a specific time scale, and this phase dependency is diminished when the causal-related intrinsic component is removed from the effect. Furthermore, we demonstrate the generic applicability of our method to both stochastic and deterministic systems, and show the consistency of the causal decomposition method compared to existing methods, and finally uncover the key mode of causal interactions in both the modelled and actual predator–prey system. We anticipate that this novel approach will assist with revealing causal interactions in complex networks not accounted for by current methods.

**Introduction**

Since the philosophical inception of causality by Galilei (1) and Hume (2) that cause must precede the effect in time, the scientific criteria for assessing causal relationships between two time series have been dominated by the notion of prediction, as proposed by Granger (3). Namely, the causal relationship from Variable A to Variable B is inferred if the history of Variable A is helpful in predicting the value of Variable B, rather than using information from the history of variable B alone.

Granger causality is based on the *time dependency* between cause and effect (4). As discussed by Sugihara *et al.* (5), Granger causality is critically dependent on the assumption that cause and effect are separable (3). While the *separability* is often satisfied in linear stochastic systems where Granger causality works well, it might not be applicable in nonlinear deterministic systems where separability appears to be impossible because both cause and effect are embedded in a nonseparable higher dimension trajectory (6, 7). Consequently, Sugihara *et al.* proposed the convergent cross mapping (CCM) method based on state-space reconstruction (5). In this context, cause and effect are *state dependent*, and Variable A is said to causally influence Variable B, although counterintuitive, if the state of Variable B can be used to predict the state of Variable A in the embedded space, and this predictability improves (i.e., converges) as the time series length increases.

Existing methods of detecting causality in time series are predominantly based on the Bayesian (8) concept of prediction. However, cause and effect are likely simultaneous (9). The succession in time of the cause and effect is produced because the cause cannot achieve the total of its effect in one moment. At the moment when the effect first manifests, it is always simultaneous with its cause. Moreover, most real-world causal interactions are reciprocal;

examples include predator–prey relationships and the physiologic regulation of body functions. In this sense, predictive causality may fail because the attempt to estimate the effect with the history of cause is compromised because the history of the cause is already simultaneously influenced by the effect itself, and vice versa.

Another drawback of the generalized prediction framework is that it requires *a priori* knowledge of the extent of past history that may influence the future, such as the time lag between cause and effect in Granger's paradigm, or the embedding dimensions in state-space reconstructions such as CCM. Furthermore, a causality assessment is incomplete if it is based exclusively on *time dependency* or *state dependency*. Time series commonly observed in nature, including those from physiologic system or spontaneous brain activity, contain oscillatory components within specific frequency bands (10, 11). The application of either linear Granger causality or the nonlinear CCM method alone is unlikely to accommodate the complex causal compositions typically observed in real-world data blended with stochastic and deterministic mechanisms.

Here, we present a causal decomposition analysis that is not based on prediction, and more importantly, is neither based on *time dependency* nor *state dependency*, but based on the *instantaneous phase dependency* between cause and effect. The causal decomposition essentially involves two assumptions: (1) any cause–effect relationship can be quantified with instantaneous phase dependency between the source and target decomposed as intrinsic components at specific time scale, and (2) the phase dynamics in the target originating from the source are separable from the target itself. We validate the method with both stochastic and deterministic systems and illustrate its application to ecological time series data of prey and predators.

**Methods**

**Causal relationship based on instantaneous phase dependency**

We defined the cause–effect relationship between Time Series A and Time Series B as follows: Variable A causes Variable B if the *instantaneous phase dependency* between A and B is diminished when the intrinsic component in B that is causally related to A is removed from B itself, but not vice versa.

$$Coh(A, B') < Coh(A, B) \sim Coh(A', B) \tag{1}$$

where *Coh* denotes the instantaneous phase dependency (i.e., coherence) between the intrinsic components of two time series, and the accent represents the time series where the intrinsic components relevant to cause effect dynamics were removed. The realisation of this definition requires two key treatments of the time series. First, the time series must be decomposed into intrinsic components to recover the cause–effect relationship at a specific time scale. Second, a phase coherence measurement was required to measure the instantaneous phase dependency between the intrinsic components decomposed from cause–effect time series.

**Empirical mode decomposition**

To achieve this, we decomposed a time series into a finite number of intrinsic mode functions (IMFs) by using the ensemble empirical mode decomposition (ensemble EMD) (12-14) technique. Ensemble EMD is an adaptive decomposition method for separating different modes of frequency and amplitude modulations in the time domain (12, 13); thus, it can be used to delineate instantaneous phase dependency in nonlinear and nonstationary data (15), thereby capturing simultaneous causal relationships not accounted for by predictive causality methods.

Moreover, the IMFs are orthogonal to each other and are separable in the statistical sense; hence solving the problems of separability in Granger's paradigm.

Briefly, the ensemble EMD (13, 14, 16) is a noise-assisted data analysis method that defines the true IMF components $S_j(t)$ as the mean of an ensemble of trials, each consisting of the signal plus white noise of a finite amplitude.

$$S_j(t) = \lim_{N \to \infty} \frac{1}{N} \sum_{k=1}^{N} \{S_j(t) + r \times w_k(t)\} \tag{2}$$

where $w_k(t)$ is the added white noise, and $k$ is the $k$th trial of the $j$th IMF in the noise-added signal. The magnitude of the added noise $r$ is critical to determining the separability of the IMFs (i.e., $r$ is a fraction of a standard deviation of the original signal). The number of trials in the ensemble $N$ must be large so that the added noise in each trial is cancelled out in the ensemble mean of large trials ($N = 1000$ in this study). The purpose of the added noise in the ensemble EMD is to provide a uniform reference frame in the time–frequency space by projecting the decomposed IMFs onto comparable scales that are independent of the nature of the original signals. With the ensemble EMD method, the intrinsic oscillations of various time scales can be separated into nonlinear and nonstationary data without any *priori* criterion on the time-frequency characteristics of the signal.

**Orthogonality and separability of IMFs**

Additionally, the orthogonality and separability of IMFs are critically dependent on selecting the magnitude $r$ of the added white noise in ensemble EMD. Because $r$ is the only parameter involved in the causal decomposition analysis, the strategy of selecting $r$ is to maximize the separability while maintaining the orthogonality of the IMFs, thereby avoiding spurious causal detection resulting from poor separation of a given signal. We calculated the nonorthogonal

leakage(12) and root-mean-square (RMS) of the pairwise correlations of the IMFs for each $r$ in the uniform space between 0.05 and 1. A general guideline for selecting $r$ in this study was to minimize the RMS of the pairwise correlations of the IMFs (ideally under 0.05) while maintaining the nonorthogonal leakage also under 0.05.

**Phase coherence**

Next, the Hilbert transform was applied to calculate the instantaneous phase of each IMF and to determine the *phase coherence* between the corresponding IMFs of two time series (17). For each corresponding pair of IMFs from the two time series, denoted as $S_{1j}(t)$ and $S_{2j}(t)$, and can be expressed as

$$S_{1j}(t) = A_{1j}(t)cos\emptyset_{1j}(t) \text{ and } S_{2j}(t) = A_{2j}(t)cos\emptyset_{2j}(t), \tag{3}$$

where $A_{1j}$, $\emptyset_{1j}$ can be calculated by applying the Hilbert transform, defined as $S_{1jH} = \frac{1}{\pi}\int\frac{S_{1j}(t')}{t-t'}dt'$, and $A_{1j}(t) = \sqrt{S_{1j}^2(t) + S_{1jH}^2(t)}$, and $\emptyset_{1j}(t) = \arctan(\frac{S_{1jH}(t)}{S_{1j}(t)})$; and similarly applied for $S_{2jH}$, $A_{2j}$, and $\emptyset_{2j}$. The instantaneous phase difference is simply expressed as $\Delta\emptyset_{12j}(t) = \emptyset_{2j}(t) - \emptyset_{1j}(t)$. If two signals are highly coherent, then the phase difference is constant; otherwise, it fluctuates considerably with time. Therefore, the instantaneous phase coherence *Coh* measurement can be defined as

$$Coh(S_{1j}, S_{2j}) = \frac{1}{T}\left|\int_0^T e^{i\Delta\emptyset_{12j}(t)}dt\right|. \tag{4}$$

Note that the integrand (i.e., $e^{i\Delta\emptyset_{12j}(t)}$) is a vector of unit length on the complex plane, pointing toward the direction which forms an angle of $\Delta\emptyset_{12j}(t)$ with the $+x$ axis. If the instantaneous phase difference varied little over the entire signal, then the phase coherence is close to 1. If the instantaneous phase difference changes markedly over the time, then the coherence is close to 0, resulting from adding a set of vectors pointing in all possible directions. This phase coherence definition allowed the instantaneous phase dependency to be calculated without being subjected to the effect of time lag between cause and effect (i.e., the time precedence principle), thus avoiding the constraints of time lag in predictive causality methods (10).

**Quantification of causal relationship between two time series**

With the decomposition of the signals by ensemble EMD and measurement of the instantaneous phase coherence between the IMFs, the most critical step in the causal decomposition analysis was the removal of an IMF followed by redecomposition of the time series (i.e., the decomposition and redecomposition procedure). If the phase dynamic of an IMF in a target time series is influenced by the source time series, removing this IMF in the target time series (i.e., subtract an IMF from the original target time series) with redecomposition into a new set of IMFs results in the redistribution of phase dynamics into the emptied space of the corresponding IMF. Furthermore, because the causal-related IMF was removed, redistribution of the phase dynamics into the corresponding IMF would be exclusively from the intrinsic dynamics of the target time series, which is irrelevant to the dynamics of the source time series, thus reducing the instantaneous phase coherence between the paired IMFs of the source time series and redecomposed target time series. By contrast, this phenomenon does not present when a corresponding IMF is removed from the source time series because the dynamics of that IMF are intrinsic to the source time series and removal of that IMF with redecomposition would still preserve the original phase dynamics from the other IMFs. Therefore, this decomposition and redecomposition procedure enables quantifying the differential causality between the corresponding IMFs of two time series.

Because each IMF represents a dynamic process operating at distinct time scales, we treated the phase coherence between the paired IMFs as the coordinates in a multidimensional space, and quantified the variance-weighted Euclidean distance between the phase coherence of the paired IMFs decomposed from the original signals as well as the paired original and redecomposed IMFs, which are expressed as follows:

$$D(S_{1j} \rightarrow S_{2j}) = \left\{ \sum_{j=1}^{m} \omega_j [Coh(S_{1j}, S_{2j}) - Coh(S_{1j}, S'_{2j})]^2 \right\}^{\frac{1}{2}}$$

$$D(S_{2j} \rightarrow S_{1j}) = \left\{ \sum_{j=1}^{m} \omega_j [Coh(S_{1j}, S_{2j}) - Coh(S'_{1j}, S_{2j})]^2 \right\}^{\frac{1}{2}} \quad (5)$$

$$\omega_j = (Var_{1j} \times Var_{2j}) / \sum_{j=1}^{m} (Var_{1j} \times Var_{2j}).$$

The relative causal strength between IMF $S_{1j}$ and $S_{2j}$ can be quantified as the relative ratio of $D(S_{1j} \rightarrow S_{2j})$ and $D(S_{2j} \rightarrow S_{2j})$, expressed as follows:

$$C(S_{1j} \rightarrow S_{2j}) = D(S_{1j} \rightarrow S_{2j}) / [D(S_{1j} \rightarrow S_{2j}) + D(S_{2j} \rightarrow S_{1j})]$$

$$C(S_{2j} \rightarrow S_{1j}) = D(S_{2j} \rightarrow S_{1j}) / [D(S_{1j} \rightarrow S_{2j}) + D(S_{2j} \rightarrow S_{1j})]. \quad (6)$$

This decomposition and redecomposition procedure was repeated for each paired IMF to obtain the relative causal strengths at each time scale, where a ratio of 0.5 indicates either that there is no causal relationship or equal causal strength in the case of reciprocal causation, and a ratio toward 1 or 0 indicates a strong differential causal influence from one time series to another. The range of D is between 0 and 1. To avoid a singularity when both $D(S_{1j} \rightarrow S_{2j})$ and $D(S_{2j} \rightarrow S_{1j})$ approach zero (i.e., no causal change in phase coherence with the redecomposition procedure), $D + 1$ was used to calculated the relative causal strength when both D values are less than 0.05.

In summary, causal decomposition comprises the following three key steps: (1) decomposition of a time series into a set of IMFs and determining the instantaneous phase coherence between each paired IMFs; (2) performing the redecomposition procedure for each paired IMFs and recalculate the instantaneous phase coherence between the paired original IMFs and redecomposed IMFs; and (3) determining the relative causal strength by estimating the deviation of phase coherence from the phase coherence of the original time series to either of the redecomposed time series.

**Validation of causal strength**

To validate the causal strength, a leave-one-sample-out cross-validation was performed for each causal decomposition test. Briefly, we deleted a time point for each leave-one-out test and obtained a distribution of causal strength for all runs where the total number of observations was less than 100, or a maximum of 100 random leave-one-out tests where the total number of time points was higher than 100. A median value of causal strength was observed.

**Deterministic and stochastic model data**

The deterministic model was used in accordance with Sugihara et al. (5) based on a coupled two-species nonlinear logistic difference system, expressed as follows (initial value $x(1) = 0.2$, and $y(1) = 0.4$):

$$x(t + 1) = x(t)[3.8 - 3.8x(t) - 0.02y(t)]$$

$$y(t + 1) = y(t)[3.5 - 3.5y(t) - 0.1x(t)] \qquad (7)$$

We also tested the logistic model proposed by Sugihara et al. for nonseparability in the state space(5), expressed as follows (initial value $x(1) = 0.2$, and $y(1) = 0.4$):

$$x(t + 1) = 3.9x(t)[1 - x(t) - \beta y(t)]$$

$$y(t + 1) = 3.7y(t)[1 - y(t) - 0.2x(t)] \qquad (8)$$

Here, we test $\beta = 0$ and $\beta > 0$ for unidirectional and bidirectional coupling, respectively.

For the stochastic model, we used part of the example shown in Ding et al. (10) for Granger causality, which is expressed as follows (using a random number as the initial value).

$$x(t + 1) = 0.95\sqrt{2}x(t) - 0.9025x(t - 1) + w_1(t)$$

$$y(t + 1) = 0.5x(t - 1) + w_2(t) \tag{9}$$

**Ecological data and validation**

We assessed the causality measures in both modelled and actual predator and prey system. The Lotka Volterra predator–prey model (18, 19) is expressed as follows:

$$\frac{dx}{dt} = \alpha x - \beta xy$$

$$\frac{dy}{dt} = \delta xy - \gamma y \tag{10}$$

where $x$ and $y$ denote the prey and the predator, respectively ($\alpha = 1$, $\beta = 0.05$, $\delta = 0.02$, $\gamma = 0.5$ were used in this study).

Experimental data on *Paramecium* and *Didinium* are available online (20), and these were obtained by scanning the graphics in Veilleux (21) and digitising the time series. Wolf and moose field data are available online at the United States Isle Royale National Park (22). The lynx and hare data were reconstructed from fur trading records obtained from Hudson's Bay Company (23). The benchmark time series (24) was reconstructed from various sources in two periods (the 1844–1904 data were reconstructed from fur records, whereas the 1905–1935 data were derived from questionnaires) (23). We used the fur-record time series between the year 1900 and 1920 for illustrative purposes.

## Results

### Illustration of the causal decomposition method

Figure 1 depicts how the causal decomposition can be used to identify the predator–prey causal relationship of *Didinium* and *Paramecium* (21). Briefly, we decomposed the time series of *Didinium* and *Paramecium* into two set of IMFs, and determined the instantaneous phase coherence (17) between comparable IMFs from the two time series (Fig. 1a). Orthogonality and separability tests were performed to determine the ensemble EMD parameter (i.e., added noise level) that minimizes the nonorthogonal leakage and root-mean-square of the correlation between the IMFs, thereby ensuring the orthogonality and separability of the IMFs (Figs. 1d and 1e). Subsequently, we removed one of the IMFs (e.g., IMF 2) from *Paramecium* (Fig. 1b; subtract IMF 2 from the original *Paramecium* signal) and redecomposed the time series. We then calculated the phase coherence between the original IMFs of *Didinium* and redecomposed IMFs of *Paramecium*. This decomposition and redecomposition procedure was repeated for IMF 2 of *Didinium* (Fig. 1c) and generalized to all IMF pairs. This procedure enabled us to examine the differential effect of removing a causal-related IMF on the redistribution of phase dynamics in cause-and-effect variables. The relative ratio of variance-weighted Euclidian distance between the phase coherence of the original IMFs (i.e., Fig. 1a) and redecomposed IMFs (i.e., Figs. 1b and 1c) is therefore an indicator of causal strength (Fig. 1f), where a ratio of 0.5 indicates either no causality is detected or no difference in causal strength in the case of reciprocal causation, and a ratio approaching 0 or 1 indicates a strong causal influence from either Variable A or Variable B, respectively.

### Application to deterministic and stochastic models

Figure 2 depicts the causal decomposition analysis in both deterministic (5) and stochastic (10) models. The IMF with a causal influence identifies the key mechanism of the model data in stochastic (Fig. 2a) and deterministic (Fig. 2b) systems. These results indicate that the causal decomposition method is suitable for separating causal interactions not only in the stochastic system, but also in the deterministic model where nonseparability is generally assumed in the state space. Furthermore, we validated and compared the causal decomposition with existing causality methods in uncorrelated white noise with varying lengths, showing the consistency of causal decomposition in a short time series and under conditions where no causal interaction should be inferred. In addition, we assessed the effect of down-sampling and temporal shift of a time series on causal decomposition and existing methods, showing that causal decomposition is less vulnerable to spurious causality due to sampling issue (3) and is independent of temporal shift which is significantly confounded with predictive causality method (25).

**Validation of causal decomposition analysis and comparison with existing methods**

We generated 10 000 pairs of uncorrelated white noise time-series observations with varying lengths (L = 10–1000) and calculated causality based on various methods (Fig. 3a). Causal decomposition exhibited a consistent pattern of causal strengths at 0.5 (the error bar denotes the standard error of causality assessment here and in the other panels), indicating that no spurious causality was detected, even in the case of the short noise time series. Causality in the CCM methods was indicated by the difference in correlations obtained from cross-mapping the embedded state space. In the case of uncorrelated white noise, the difference of correlation should be approximately zero, indicating no causality. However, the CCM method detects spurious causality with differences of up to 0.4 in the correlations in the short time series, and the difference between the correlations decreased as the signal length increased. A high percentage or intensity of spurious causality was also observed in Granger's causality and in mutual information from the mixed embedding (MIME) method (26).

Next, we assessed the effect of down-sampling on the various causality methods (Fig. 3b). The stochastic and deterministic models shown in Fig. 2 are used (the corresponding colour for each variable is shown in the figure). The time series were down-sampled by a factor 1 to 10. For Factor 1, the time series were identical to the original signals. The down-sampling procedure destroyed the causal dynamics in both models and made causal inference difficult in predictive causality analysis (25). Causal decomposition analysis revealed a consistent pattern absent of causality when the causal dynamics were destroyed when the down-sampling factor was greater than 2. However, spurious causality was detected with the predictive causality methods when the signals were down-sampled.

Finally, we evaluated the effect of temporal shift on the causality measures (Fig. 3c). Temporal shift (both lagged or advanced up to 20 data points) was applied to both the stochastic and deterministic time series. Causal decomposition exhibited a stable pattern of causal strength independent of a temporal shift up to 20 data points. As anticipated, the predictive causality methods lost their predictability even when the corresponding time points were shifted only slightly.

**Quantifying predator-prey relationship**

Figure 4 shows the results of applying causal decomposition to ecosystem data from the Lotka Volterra predator–prey model (18, 19) (Fig. 4a), wolf and moose data from Isle Royale National Park (22) (Fig. 4b), and the Canada lynx and snowshoe hare time series reconstructed from historical fur records of Hudson's Bay Company (23) (Fig. 4c). The causal decomposition invariantly identifies the dominant causal role of the predator in the IMF, which is consistent with the classic Lotka Volterra predator–prey model. Previously, the causality of such autonomous differential equation models is understood only in mathematical terms because there is no prediction-based causal factor (27), yet our results indicated that the causal influence of this model can be established through the decomposition of instantaneous phase dependency.

**Comparison of causal assessment in ecosystem data with existing methods.**

Fig. 5 showed the comparison of causality assessment in these predator and prey data using different methods. In general, results showed that neither the Granger nor CCM methods consistently identify predator–prey interactions in these data, indicating that the predator–prey relationship does not exclusively fit either the stochastic or deterministic chaos paradigms. CCM method could not be used to detect causality in the Lotka Volterra predator–prey model, and it exhibited a cross-over of correlations in the wolf and moose data. The CCM result was consistent with the data presented by Suigihara et al. (5) based on in vitro experiment of *Didinium* and *Paramecium* interactions (21). Granger's causality generally detects mutual interactions, but the causal strengths (F-test) were inconsistent in ecosystem data (the vertical dashed line denotes the significance threshold with $P < 0.05$), which the findings are also consistent with the supplementary data in Sugihara et al.(5); similarly, the inconsistency in causal strength was also observed in the results obtained with the MIME method.

**Discussion**

An interdisciplinary problem of detecting causal interactions between oscillatory systems solely from their output time series has attracted considerable attention for a long time. The motivation of causal decomposition analysis is that the inference of causality should not be largely dependent on the temporal precedence principle. In other words, observing the past with a limited period is insufficient to infer causality because that history is already biased. Instead, we followed the most fundamental criterion of causal assessment proposed by Galilei (1): "cause is that which put, the effect follows; and removed, the effect is removed." Therefore, the complex dynamical process between cause and effect should be delineated through the decomposition of intrinsic causal components inherited in causal interactions. Our approach does not neglect the temporal precedence principle, but emphasise the instantaneous relationship of causal interaction, and is thus more generic in terms of detecting simultaneous or reciprocal causation, which is not fully accounted for by predictive methods.

Because our causal strengths measurement is relative, it detects differential causality rather than absolute causality. Differential causality adds to the philosophical concept of mutual causality that all causal effects are not equal, and it may fit the emerging research data better than linear and unidirectional causal theories do. In addition, causal decomposition using EMD fundamentally differs from the spectral extension of Granger's causality (28) in that the latter involves the prior knowledge of history (e.g., autoregressive model order) and is susceptible to nonstationary artefacts. Furthermore, without resorting to frequency-domain decomposition, EMD bypass the uncertainty principle imposed on data characteristics as in Fourier analysis, and results in more precise phase and amplitude definition (29).

The operational definition of causal decomposition is in accordance with Granger's assumption on separability (3) but in a more complete form. We note that such definition is distinct with non-separability assumed by CCM. Clearly, CCM is developed under the constraints of perfect deterministic system, in which the state of cause is encoded in effect that is not separable from effect itself. The state-space reconstruction approach such as CCM may be applicable to certain ecosystem data, such as predator and prey interactions, in which they represent non-separable components of the ecosystem (30), but is unlikely to generalize to all causal interactions being studied (31). The use of EMD overcomes the difficulty of signal decomposition in nonlinear and nonstationary data, and is thus applicable to both stochastic and deterministic systems in that the intrinsic components in the latter remain separable in the time domain. Furthermore, the central element in causal decomposition analysis is the decomposition and redecomposition procedure, and we do not exclude the use of other signal decomposition methods (32) to detect causality in a similar manner. Therefore, the development of causal decomposition is not to complement existing methods, but to explore a new territory of time series analysis for assessing causality. We anticipate that this novel approach will assist with revealing causal interactions in complex networks not accounted for by current methods.

**Acknowledgments** This work was supported by the Ministry of Science and Technology (MOST) of Taiwan (grant MOST 101-2314-B-075 -041-MY3; 104-2314-B-075 -078 -MY2); and the MOST support for the Center for Dynamical Biomarkers and Translational Medicine, National Central University, Taiwan (grant MOST 103-2911-I-008-001).

**Author Information** The authors declare no competing financial interests.

**Figure Legends**

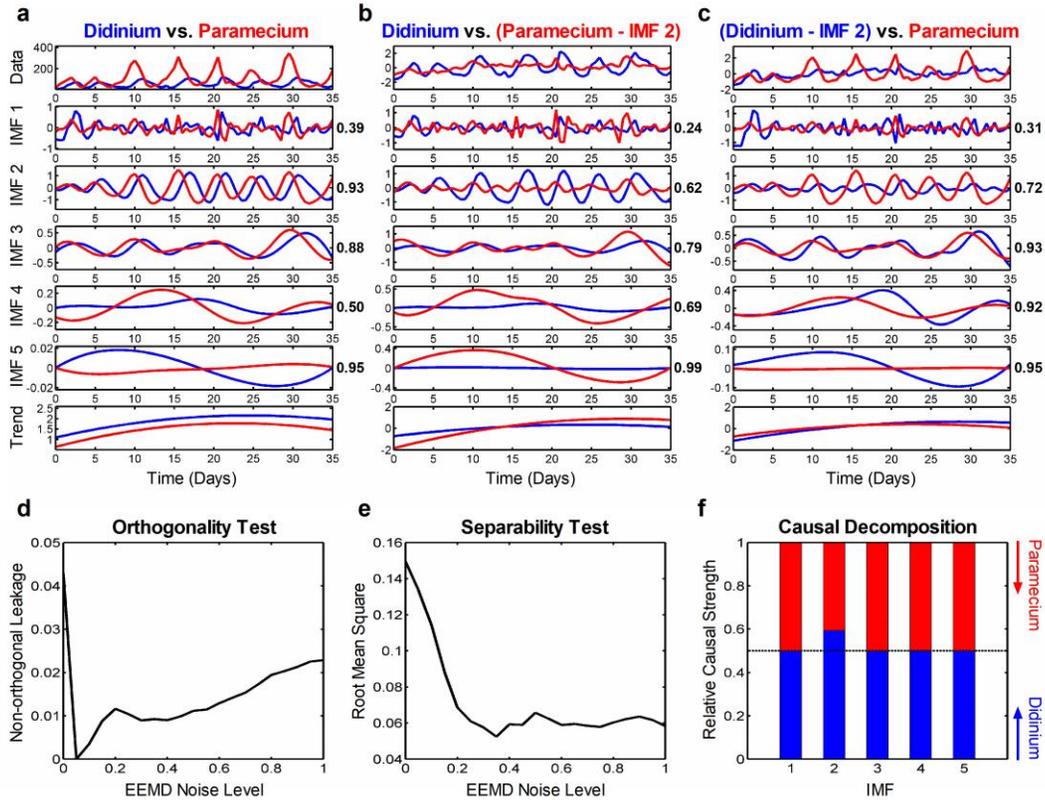

**Fig. 1 Causal decomposition analysis. a**, Ensemble EMD analysis of *Didinium* and *Paramecium* time series yields five IMFs (i.e., stationary components) and a residual trend (i.e., nonstationary trend). Each IMF operated at distinct time scales. Phase coherence values between comparable IMFs are shown at the right side of the panel. **b,** Removal of an IMF (e.g., IMF 2) from *Paramecium* with redecomposition leads to a declined phase coherence between the original *Didinium* IMFs and redecomposed *Paramecium* IMFs. **c,** Repeating the same procedure in the *Didinium* time series resulted in a smaller decrease in phase coherence between the redecomposed *Didinium* IMFs and the original *Paramecium* IMFs. The causal strengths between *Didinium* and *Paramecium* can be estimated by the relative ratio of variance-weighted Euclidian distance of the phase coherence between **b** and **a** (for *Didinium*), and between **c** and **a** (for *Paramecium*). The ability of ensemble EMD to separate time series depends on the orthogonality

and separability of the IMFs with added noise, which can be evaluated by **d,** nonorthogonal

leakages and **e,** the root-mean-square of correlations between pairwise IMFs. The strategy of

choosing the added noise level in the ensemble EMD is to maximise the separability (minimize

the root-mean-square of pairwise correlation values among IMFs < 0.05) while maintaining

acceptable nonorthogonal leakages (< 0.05). A noise level $r$ at 0.35 standard deviations of the

time series was used in this case. **f,** Generalization of causal decomposition to each IMF

uncovers a causal relationship from *Didinium* to *Paramecium* in IMF 2 but not in the other IMFs,

indicating that a time-scale-dependent causal interaction in the predator–prey system.

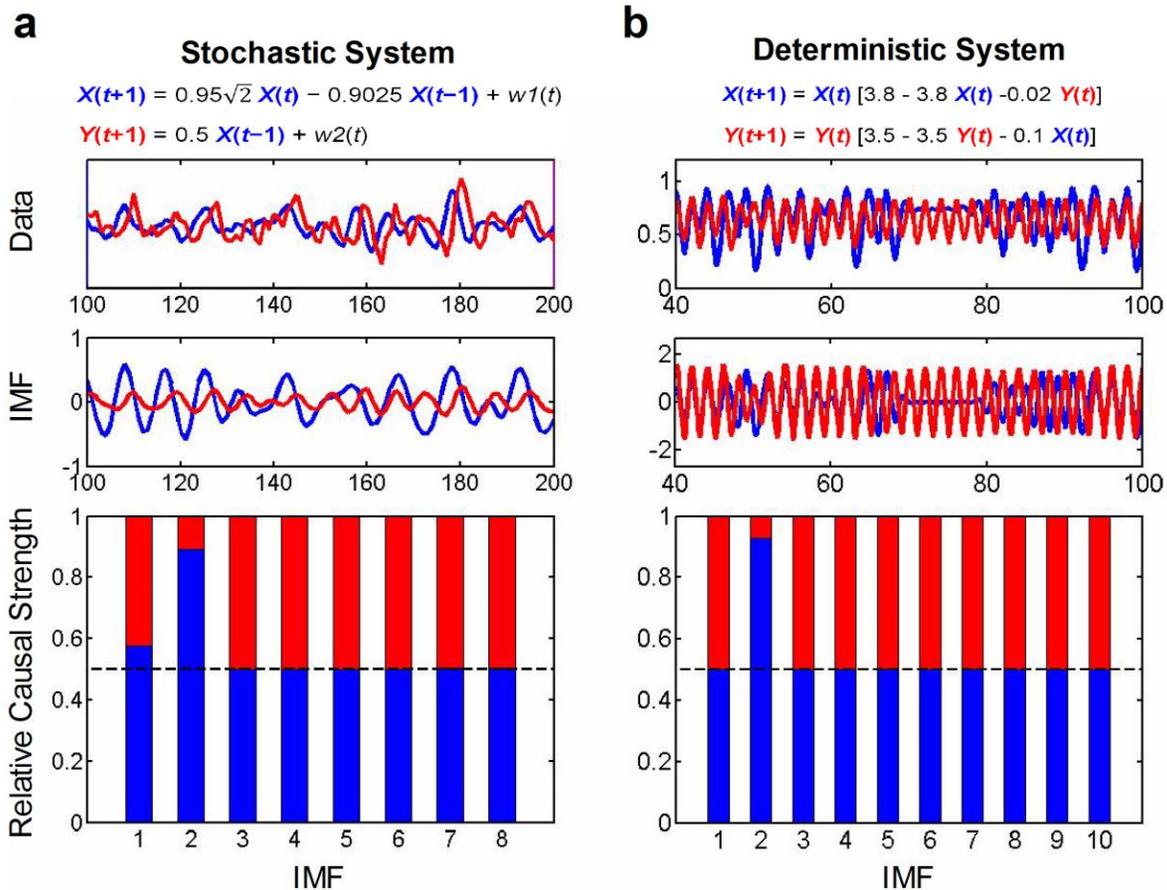

**Fig. 2 Stochastic and deterministic model evaluation.** Application of causal decomposition to **a,** stochastic system(10) and **b,** deterministic system(5) (ensemble EMD parameter $r = 0.15$ for both cases). A causal influence was identified in IMF 2 in both systems, capturing the main mode of signal dynamics in each system (e.g., a lag order of 2 between the IMFs in **a,** and chaotic behaviour of the logistic model in **b**). The causal decomposition is not only able to handle noisy data in the stochastic model, but is can also identify causal components in the deterministic model with the aid of ensemble EMD in separating weakly coupled chaotic signals into identifiable IMFs. Data lengths: **a,** 1000 data points; **b,** 400 data points.

**a**

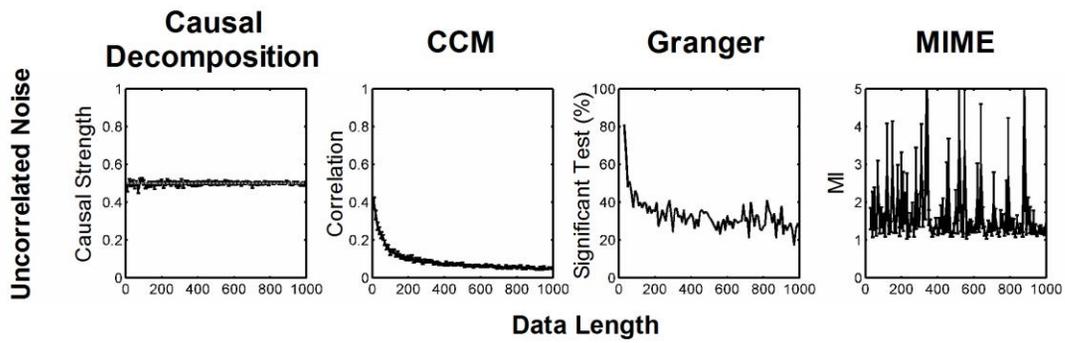

**b**

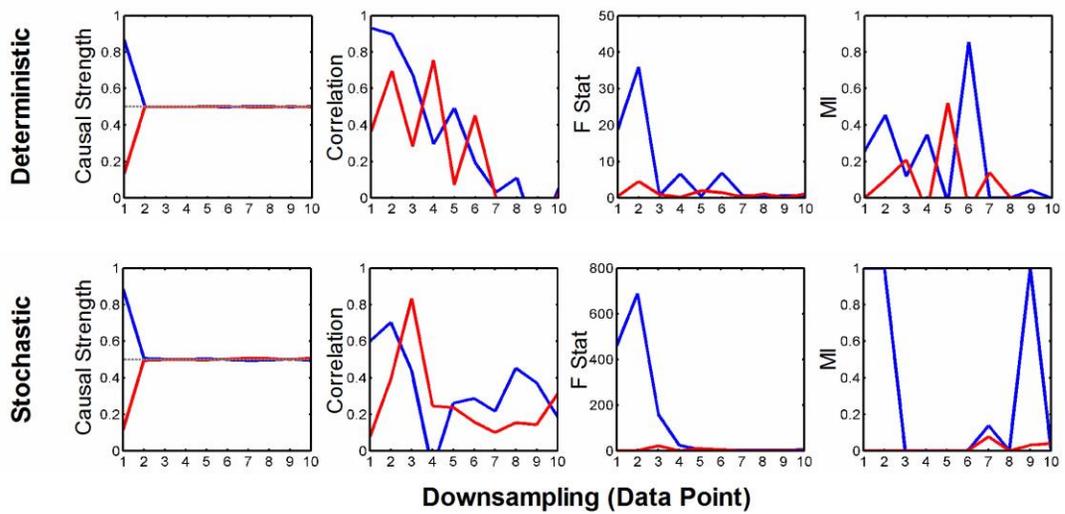

**c**

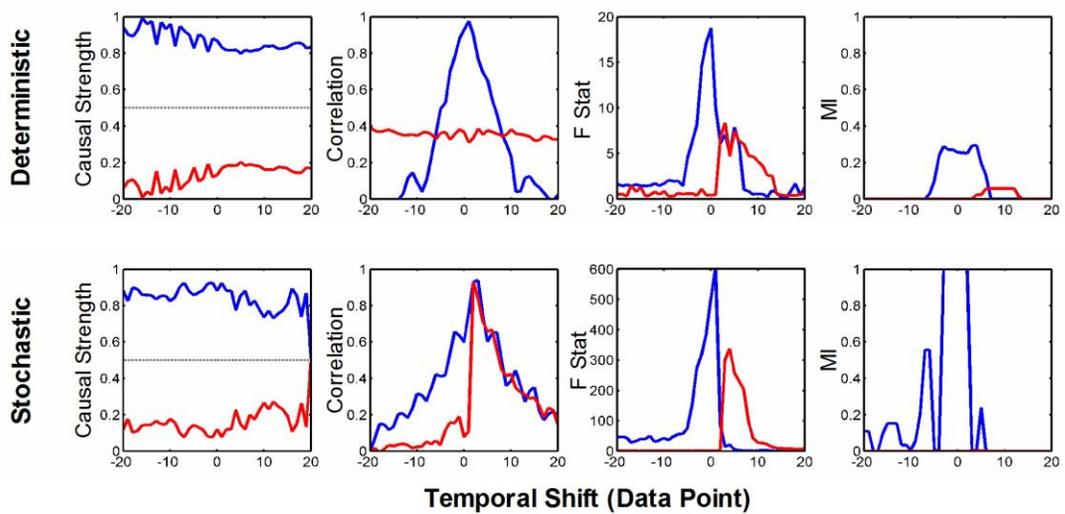

**Fig. 3 Validation of causal decomposition method a.** The finite length effect on causality assessment. **b.** Effect of down-sampling on the various causality methods. **c.** Effect of temporal shift on the various causality methods.

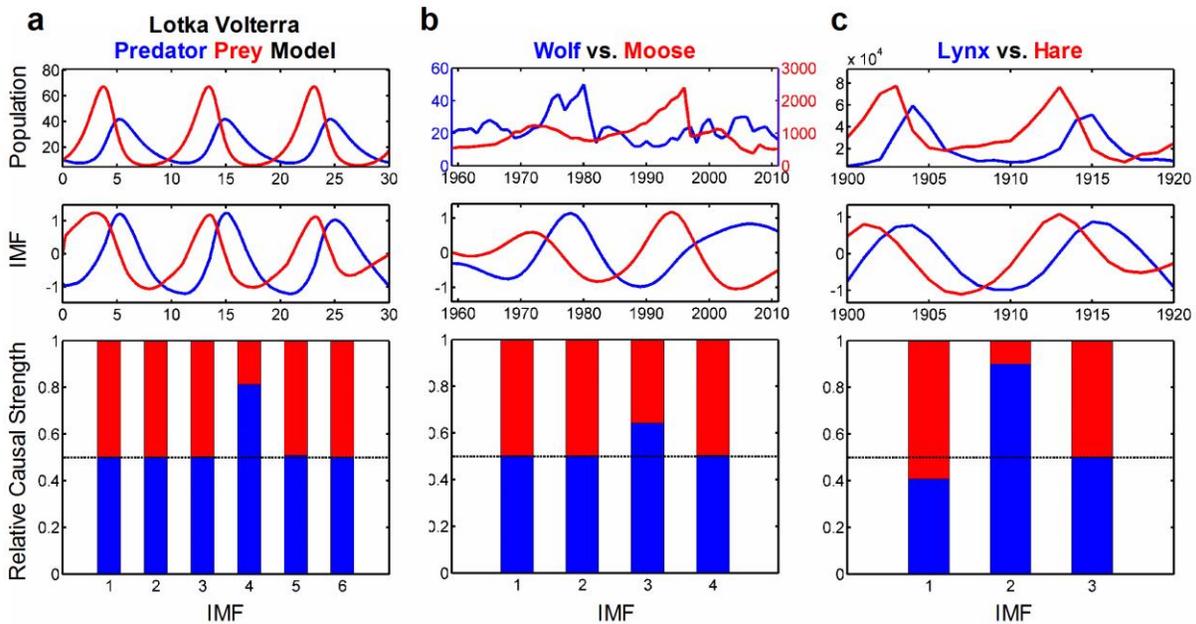

**Fig. 4 Causal decomposition of predator–prey data. a,** Lotka Volterra predator–prey model. **b,** wolf and moose time series from Isle Royale National Park in Michigan, USA. **c,** Canada lynx and snowshoe hare time series reconstructed from historical fur records of Hudson's Bay Company(23). The IMFs shown in the figure correspond to significant causal interactions identified in each observation (**a,** IMF 4, **b,** IMF 3, **c,** IMF 2). Ensemble EMD parameter: **a,** $r = 0.4$, **b,** $r = 0.3$, **c,** $r = 0.3$).

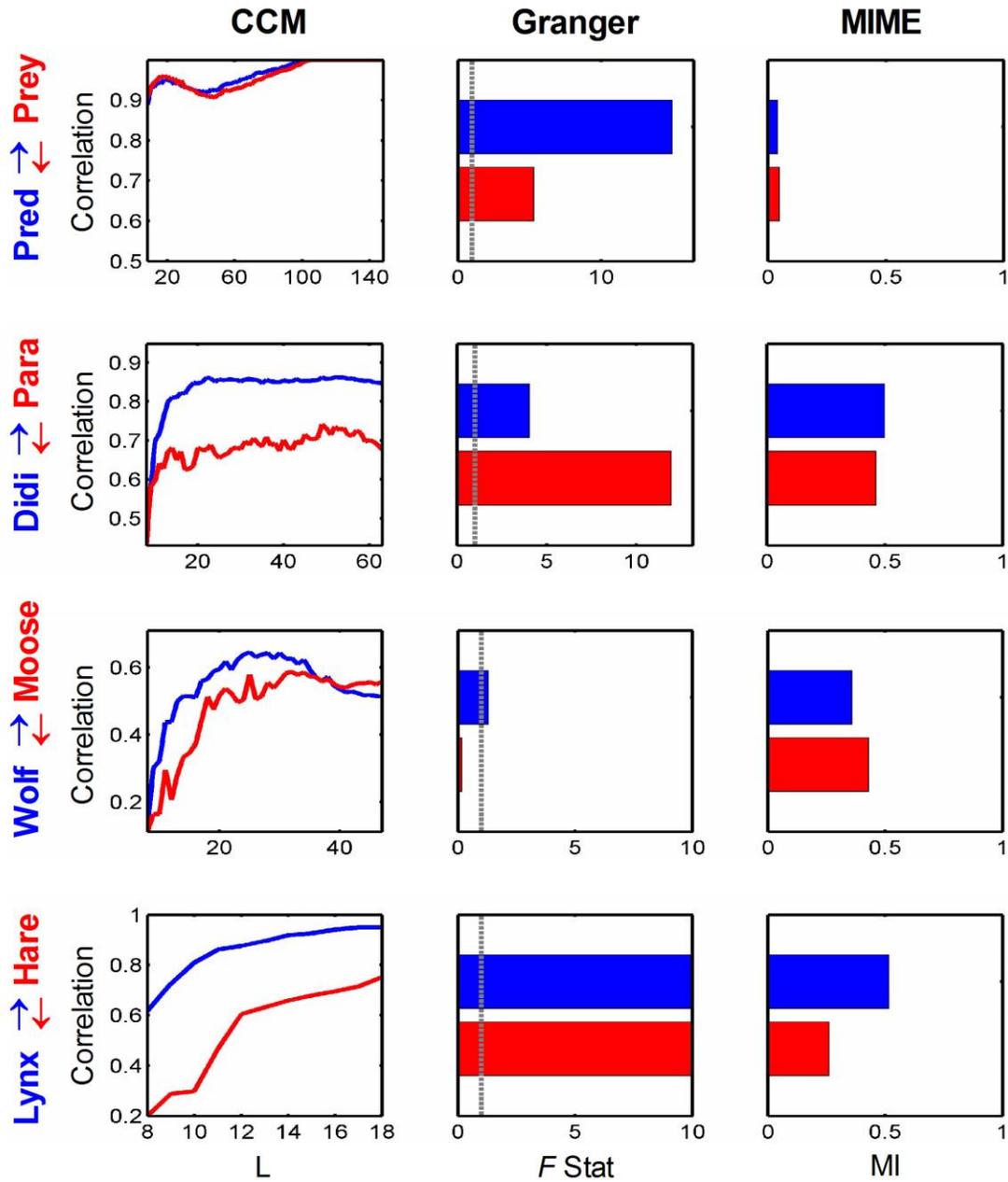

**Fig. 5.** Causal assessment in ecosystem data with existing methods.